\documentclass{appolb}
\usepackage{graphicx}

\usepackage{amsthm}
\usepackage{amsmath}
\usepackage{subfigure}
\usepackage{color}
\usepackage{listings}

\begin{document}
\title{Random sequential adsorption of unoriented cuboids with~a~square base and a comparison of cuboid-cuboid intersection tests%
\thanks{Presented at 30th M.Smoluchowski Symposium on Statistical Physics, September 3-8, 2017.}%
}
\author{Piotr Kubala, Micha\l{} Cie\'sla
\address{M.\ Smoluchowski Institute of Physics, Department of Statistical Physics, Jagiellonian University, \L{}ojasiewicza 11, 30-348 Krak\'ow, Poland.}
}
\maketitle
\begin{abstract}
In the paper, packings built of identical cuboids with a square base created by random sequential adsorption are studied. The result of the study show that the packing of the highest density are obtained for oblate and prolate cuboids of the edge-edge length ratios of $0.7$ and $1.4$. For both cases, the packing fraction is $0.400 \pm 0.002$, which is approximately 8\% higher than the value reported for cubes. Additionally, because the crucial part of the packing generation algorithm is the cuboid-cuboid intersection detection, several methods were tested. It appears that the fastest one is based on the separating axis theorem.
\end{abstract}
  
\section{Introduction}
Random sequential adsorption (RSA) \cite{Feder1980, Evans1993} is a protocol for generating random packings of arbitrary objects. It is based on subsequent repetitions of the following steps:
\begin{description}
\item[--]{A virtual, randomly oriented particle is placed inside a packing at random position.}
\item[--]{The virtual particle is checked if it intersects with other particles that were previously added to the packing.}
\item[--]{If there are no overlaps, the virtual particle is added to the packing. Otherwise it is removed and abandoned.}
\end{description}
Iterations end when there is no place for adding another particle to the packing. Such packing is called saturated.

Although RSA is most commonly used for modelling two-dimensional monolayers created in irreversible adsorption experiments, it can be also utilized to study various properties of random packings in other dimensions, e.g. \cite{Sherwood1997, Torquato2006, Zhang2013, Ciesla2018}. RSA of cubic particles became a subject of scientific investigations after Palasti came to conclusion that $\theta_d = (\theta_1)^d$, where $\theta_d$ is the mean packing fraction of d-dimensional packing of oriented cubes \cite{Palasti1960}, and $\theta_1 = 0.7476...$ was calculated analytically by by Renyi \cite{Renyi1958}. In the end, the conclusion turned out to be false \cite{Brosilow1991}; but it remains quite a good approximation of the real, saturated packing fraction \cite{Bonnier2001}. 

Random packing of unoriented cubes was studied mainly in terms of random close packings, where neighbouring particles are in touch with each other. However, properties of such packings are very sensitive to the details of experimental or numerical protocol used for packing generation \cite{Torquato2000, Asencio2017}. In 2010, Baker and Kudrolli examined packings obtained by throwing cubes at random places, similarly as by RSA, but under the influence of gravity, and obtained packing density about $0.57$. \cite{Baker2010}. Argawal et al. studied phase diagram of a packing of cubes and found that orientationally oriented phase appears when packing fraction exceeds $0.5$ \cite{Agarwal2011}. Recent study of RSA of unoriented cubes reports that the mean packing fraction is $0.3686 \pm 0.0015$ and no global orientational order is observed \cite{Ciesla2018}.

This study is focused on RSA of cuboids with a square base. The main goal is to establish how  shape anisotropy influences the mean, saturated packing fraction of cuboids. Available results for two-dimensional shapes like rectangles \cite{Vigil1989}, dimers \cite{Ciesla2014, Ciesla2015}, spherocyllinders and ellipses \cite{Ciesla2016} show that slight anisotropy of packed particles leads to denser packings. The same effect is observed for random close packing of ellipsoids \cite{Donev2004, Man2005}. Similar study of RSA packing built of uniaxial ellipsoids reports that packing fraction maxima are equal to $0.406$ and $0.411$ and are obtained for axis ratio of $0.7$ and $1.5$ \cite{Sherwood1997}. 

The secondary goal of this study is to find the most effective method for determining cuboid-cuboid intersection, which is crucial not only for efficient packing generation, but also in collision detection \cite{lin1996,koziara2005}.
\section{Model}
The packing is built of identical cuboids of edges $a^{2/3}, a^{-1/3}, a^{-1/3}$, where $a \in [0.3, 2.0]$. Note that for such cuboids the edge-edge length ratio is $a$ and the volume $V_c=1$. Cuboids are put inside a cube of volume $V=10^4$ according to RSA algorithm. Position of a cuboid centre is given by a random point selected uniformly inside the packing. The cuboid orientation is determined using three independent random numbers $(u_1, u_2, u_3)$, which are selected uniformly from $[0, 1)$ interval. The  cuboid is rotated around $x$-axis by $2\pi u_1$ radians, then around $y$-axis by $\arcsin(2u_2-1)$ radians, and then around $z$-axis by $2\pi u_3$ radians. The same method of choosing random position and orientation for a three dimensional object was used in the study of RSA of cubes \cite{Ciesla2018}. We assumed periodic boundary conditions for the whole packing. Packing generation was stopped after $n = 10^9$ iterations which corresponds to dimensionless time $t = n V_c / V = 10^5$.
For each value of parameter $a$, $100$ independent random packings were generated.

During packing generation, the mean packing fraction dependence on the number of iterations was measured:
\begin{equation}
  \theta(t) = \frac{N(t) V_c}{V},
\label{theta}
\end{equation}
where $N(t)$ is the mean number of particles in a packing after the number of iterations corresponding to the dimensionless time $t$.
\section{Overlap detection}
The crucial point of the RSA algorithm is testing whether there is an intersection between a trial object and already existing particles in a packing. In case of packing built of cuboids, this test is the most time consuming part of the RSA algorithm. There are many general intersection test for convex shapes, both numerical and analytical. One of the widely used numerical tests is an overlap potential method \cite{donev2006}, but it can be applied only for particles with smooth surfaces. This section presents two popular general-purpose overlap tests for any polyhedra and a custom test dedicated for unoriented cuboids.

\subsection{Separating axis theorem}

\begin{figure}[htb]
	\centering
	\subfigure[]{\includegraphics[width=0.45\columnwidth]{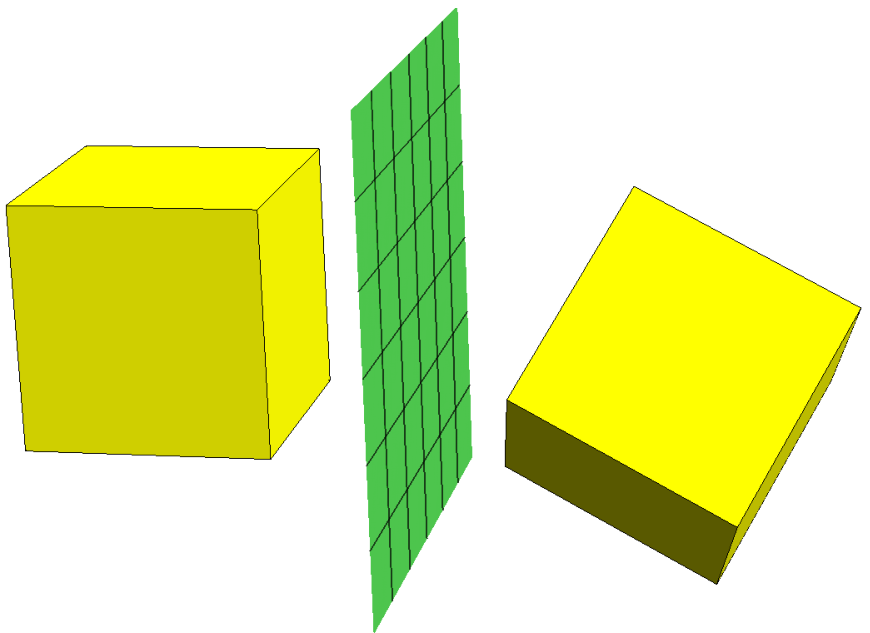}}
	\subfigure[]{\includegraphics[width=0.45\columnwidth]{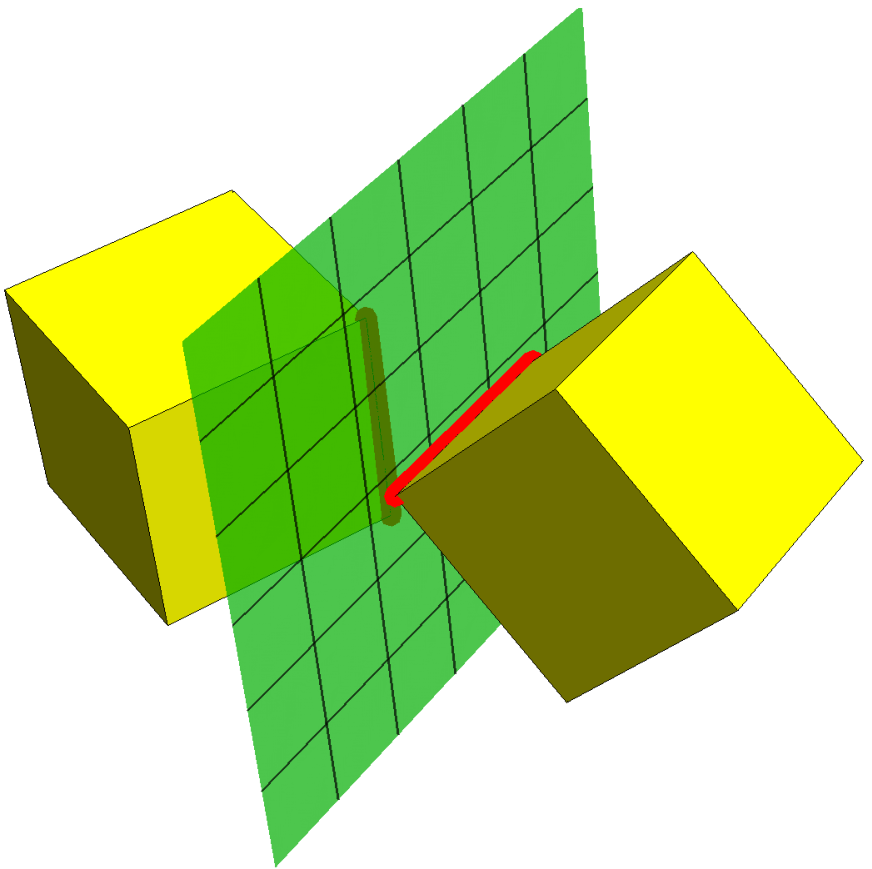}}
    \caption{Illustration of separating axis theorem. Separating plane could be parallel to one of cuboids' faces (a) or spanned by cuboids' edges (b).}
    \label{fig:sat}
\end{figure}

Separating axis theorem (SAT) states that two convex shapes are disjunctive if and only if there exists a so called separating plane, such that both objects are on different sides of this plane (see Fig.\ref{fig:sat}).
SAT can also be used for concave shapes by splitting them into convex parts and testing each pair of them. It is worth to mention that SAT can be generalized to other dimensions. In two dimensions a separating plane is a line while for higher dimension it is a hyperplane. An axis perpendicular to separating (hyper)plane is called separating axis (SA), hence the theorem name. The projections of shapes onto SA are disjunctive if and only if the shapes are disjunctive. SAT itself, unfortunately, does not point any algorithm for finding SA. However, in case of convex polyhedron the following conditions apply \cite{Huynh2008}:
\begin{itemize}
\item[--] Two convex polygons are disjunctive if and only if one of all axes perpendicular to one of either shape's edges is SA.
\item[--] Two convex polyhedra are disjunctive if and only if one of all axes perpendicular to one of either shape's faces, or perpendicular to two edges, one from each shape, is SA.
\end{itemize}
Therefore, for cuboids, the intersection test based on SAT contains the following steps:
\begin{enumerate}
\item[--] Examine all axes of cuboids. If any of them is SA, cuboids are disjunctive.
\item[--] Examine all cross products of normal vectors of two cuboid faces; one from each cuboid. If axis spanned by any of them is SA, cuboids are disjunctive.
\item[--] If none of examined axes is SA, cuboids overlap.
\end{enumerate}
Here, the advantage was taken of the fact that a cuboid's axes are parallel to lines spanned by its edges. In the study, two variants of this overlap criterion were examined. In the first one, referred as SAT test, cuboids are simply projected on a potential SA and it is tested whether the projections are disjunctive. In the second one, referred as optimized SAT test, all the optimizations described in Ref. \cite{Huynh2008} are used.

\subsection{Triangle-triangle intersection}

Triangle-triangle (TT) intersection test is based on the fact that the surface of any polyhedron can be decomposed into a set of triangles. Therefore, intersection test can be performed as follows:
\begin{enumerate}
\item[--] Divide both cuboids into triangles.
\item[--] Examine all pairs of triangles, one from each cuboid, for overlapping. If any of them overlap, so do cuboids. Otherwise, cuboids are disjunctive.
\end{enumerate}
There are many fast TT intersection tests. Here, the one described in Ref. \cite{Moller1997} was used. The test works for both, convex and concave shapes, but it does not detect if one object is fully inside the second one. However, it is not a drawback in the case of identical particles.

\subsection{Custom test}
\begin{figure}[htb]
	\centering
	\subfigure[]{\includegraphics[width=0.30\columnwidth]{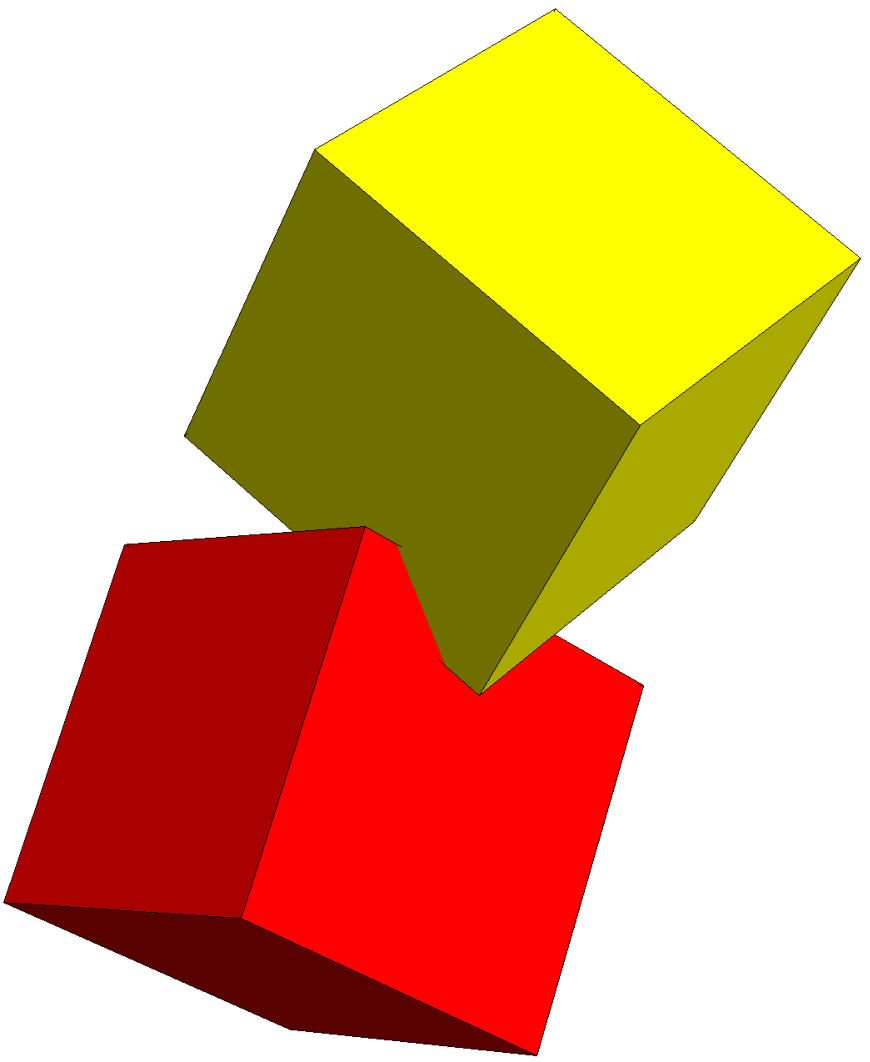}}
	\subfigure[]{\includegraphics[width=0.30\columnwidth]{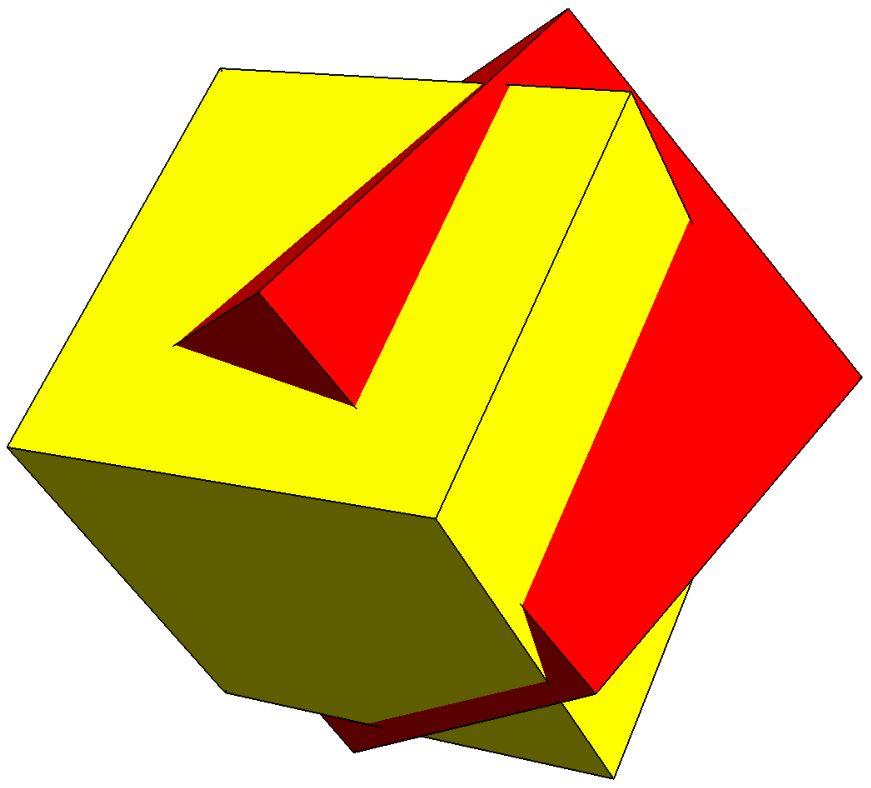}}
	\subfigure[]{\includegraphics[width=0.30\columnwidth]{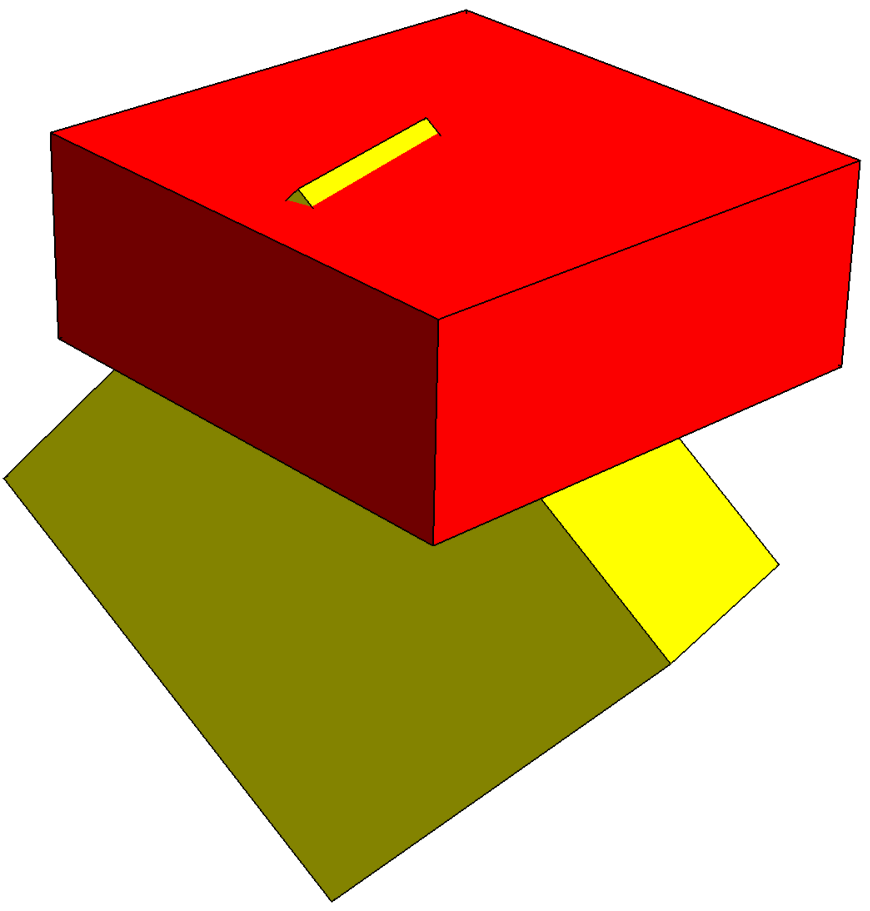}}
    \caption{Intersecting cuboids that fulfil the first and second condition of the custom test.  Additionally, panel (c) shows example configuration where only one of the two last conditions is violated.}
    \label{fig:intersections}
\end{figure}

The last test described here is valid only for cuboids and was used previously in the study of RSA of cubes \cite{Ciesla2018}. Cuboids A and B are disjunctive it and only if:
\begin{enumerate}
\item[--] Vertices of B lie outside A.
\item[--] Vertices of A lie outside B.
\item[--] Edges of B do not intersect with A's faces.
\item[--] Edges of A do not intersect with B's faces.
\end{enumerate}
The first two conditions are necessary but insufficient to ensure the lack of overlap. Fig. \ref{fig:intersections} shows some possible configurations where these conditions are fulfilled however cuboids intersect. 
Note, that for some specific edge-edge lengths ratios  the last condition can be skipped, as it results from the third one \cite{Ciesla2018}, but in general it is relevant (see Fig.\ref{fig:intersections} c).

For efficiency, it is worth to centre and align one of tested cuboids with axes of the coordinate system. 


\subsection{Speed test}

\begin{figure}[htb]
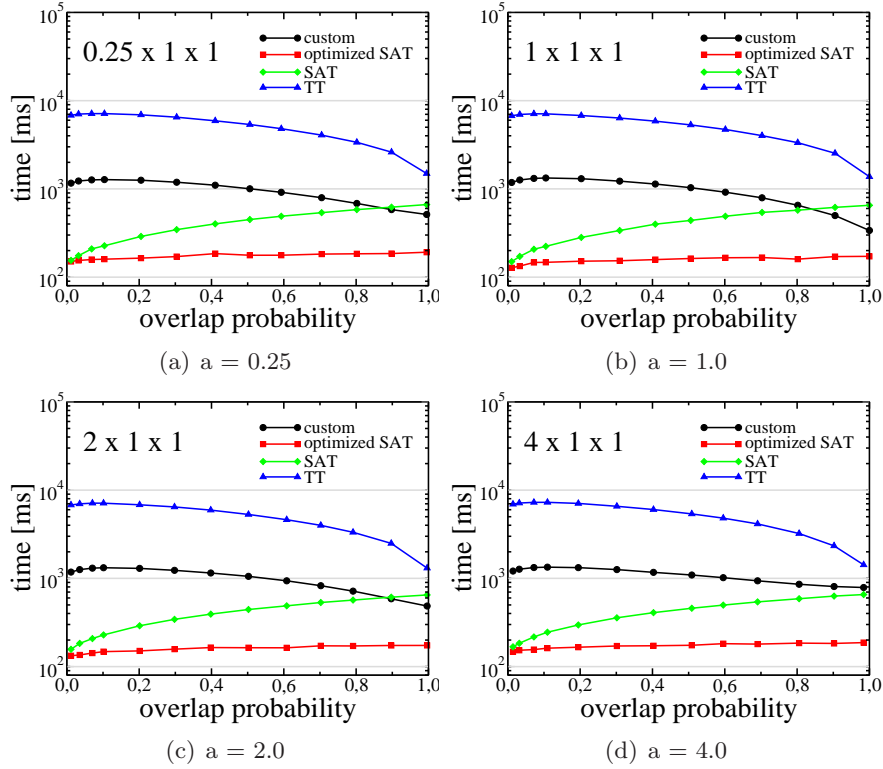

	\centering
	\subfigure[a = 0.25]{\includegraphics[width=0.45\columnwidth]{025}}
	\subfigure[a = 1.0]{\includegraphics[width=0.45\columnwidth]{1}}
	\subfigure[a = 2.0]{\includegraphics[width=0.45\columnwidth]{2}}
	\subfigure[a = 4.0]{\includegraphics[width=0.45\columnwidth]{4}}
    \caption{Dependence of the mean time of $10^6$ overlap tests on intersection probability for the studied algorithms and for different values of parameter $a$.}
    \label{fig:cube_st}
\end{figure}

It is worth noting that some algorithms, like TT test and custom test, are more efficient when overlap probability is high, while other, like SAT, run faster when most particles are disjunctive. Therefore, the choice of the best algorithm may vary depending on packing density. At the initial phase, when the packing is almost empty, intersection between an added particle and its closest neighbour in the packing is rare. On the other hand, when the packing is almost saturated, the probability of overlap is close to $1$. Therefore, it is relevant to check how the efficiency of the studied algorithms depends on overlap probability. To perform tests, centres of two cuboids were selected randomly from the ball of a given radius and their orientations were set as described in the Model section. Then overlap checks were performed. This procedure was repeated $10^6$ times for each studied intersection test. The mean time of such 10 independent experiments for several different cuboids is presented in Fig.\ref{fig:cube_st}.

The most efficient test for the whole range of intersection probability is the optimised SAT. It's implementation is given in the Appendix. In contrast with non optimised version, its speed only slightly decreases with overlap probability. Custom test is slower than SAT tests for low probabilities and its speed differs for different cuboids -- for polyhedrons of higher anisotropy the first two conditions are often enough to detect an overlap. For cubes and overlap probability greater than 0.8, it is faster than non-optimised SAT. Surprisingly, commonly used intersection criterion -- the TT test -- is up to 2 orders of magnitude slower than other tests in spite of using fast triangle-triangle intersection test.
\section{Results}
\begin{figure}[htb]
\centerline{%
\subfigure[]{\includegraphics[width=0.35\columnwidth]{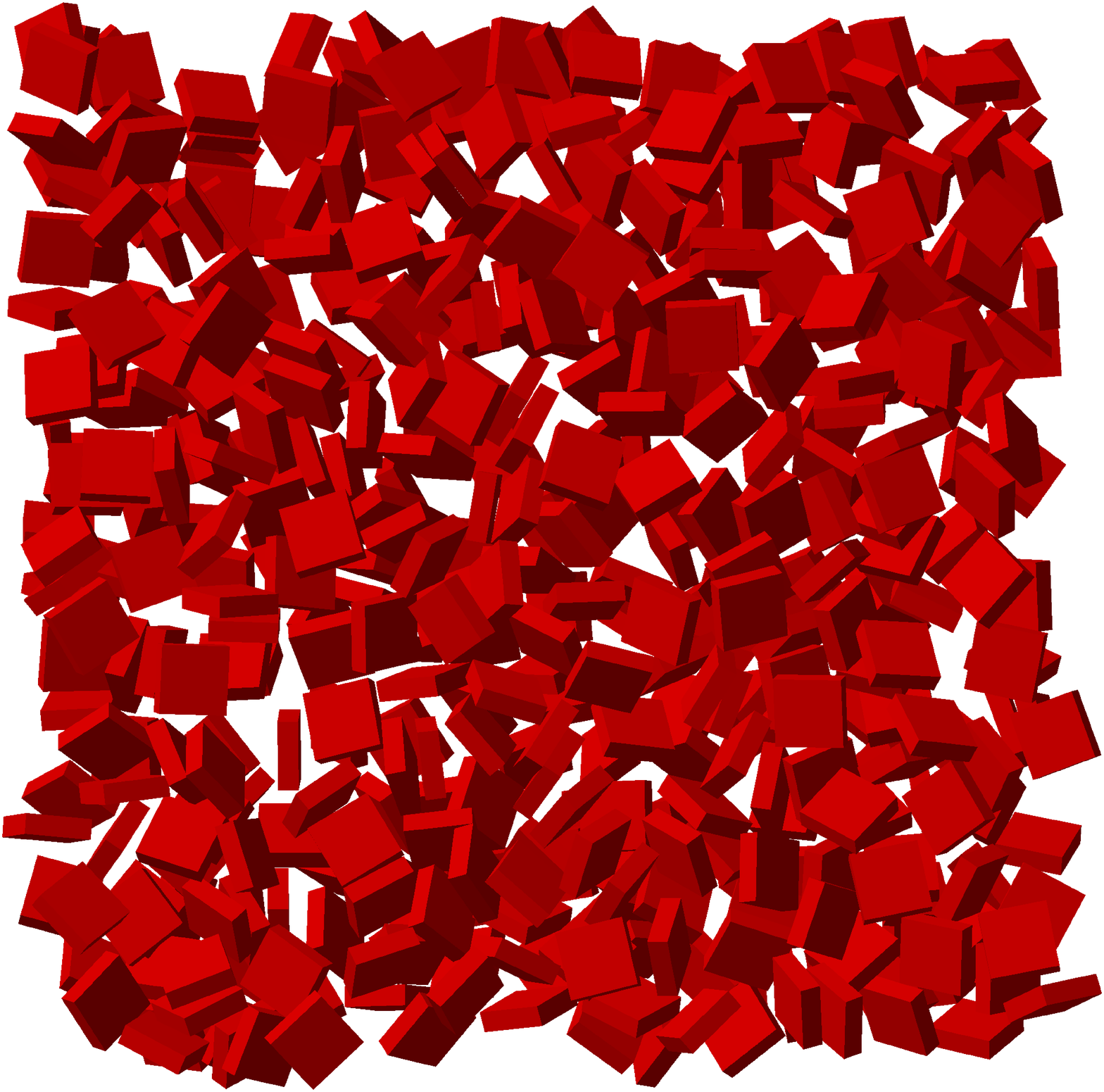}}
\subfigure[]{\includegraphics[width=0.35\columnwidth]{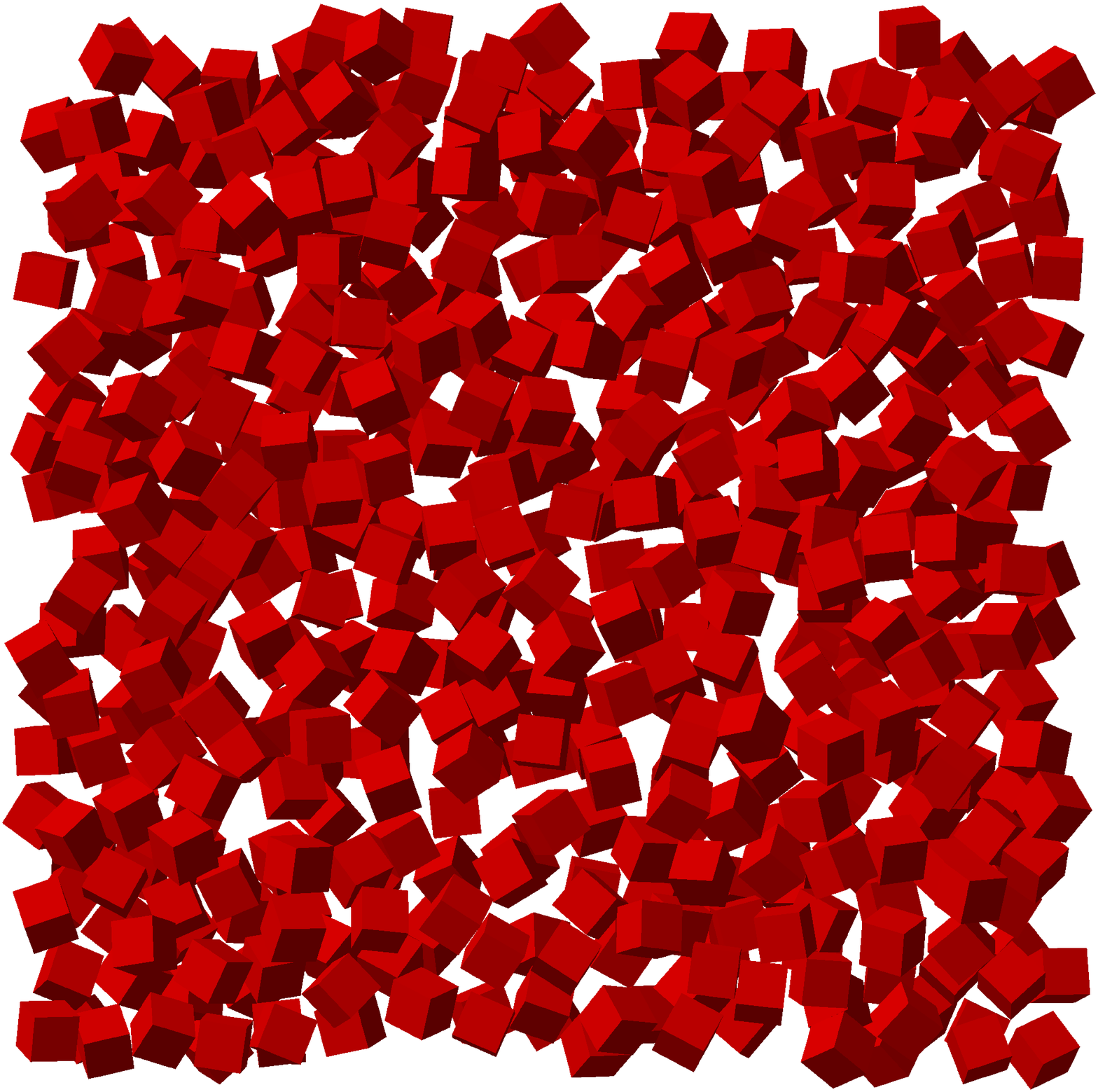}}
\subfigure[]{\includegraphics[width=0.35\columnwidth]{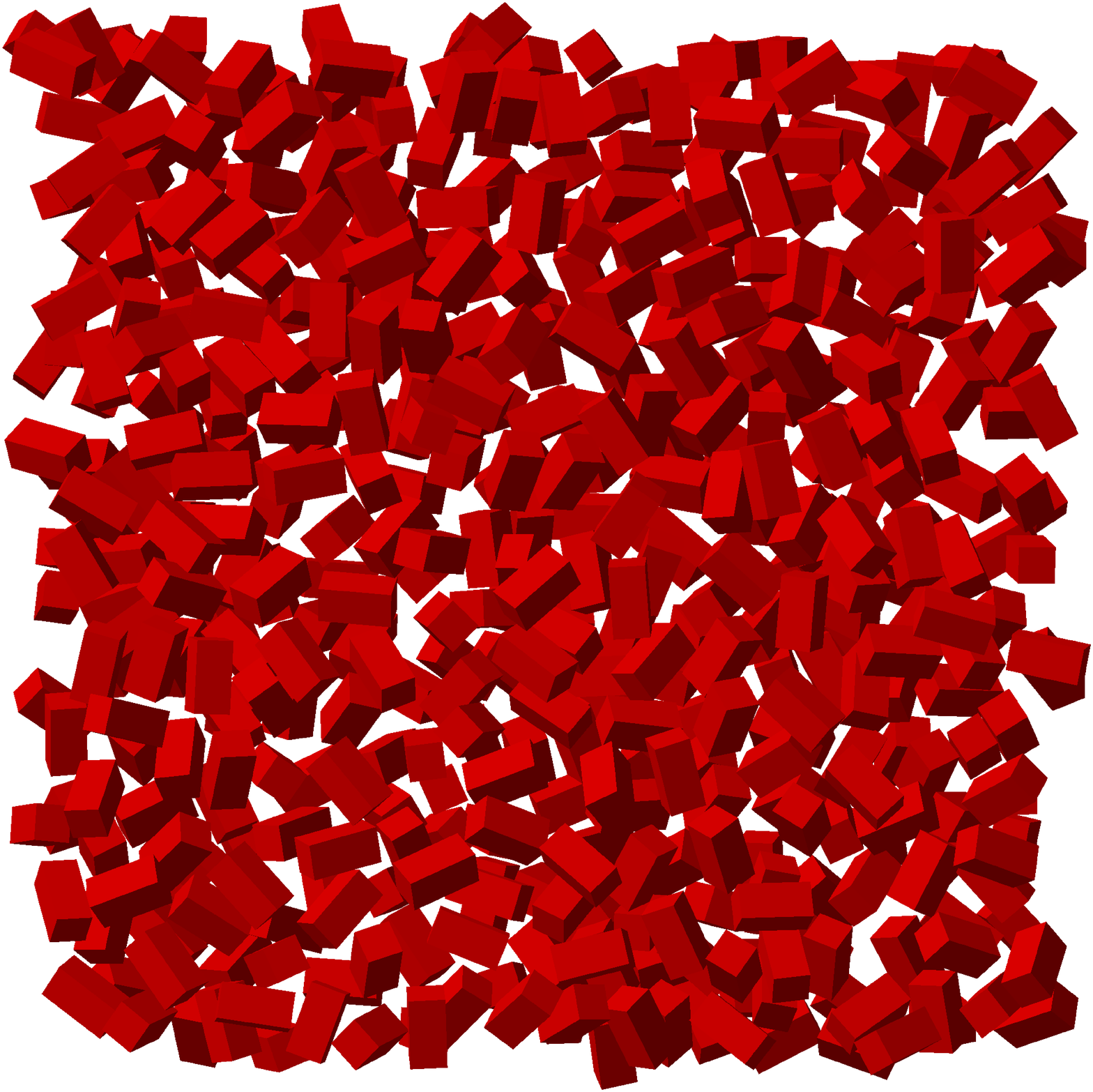}}
}
\caption{Fragments of cuboids packings generated by RSA algorithm. Packing (a) contains cuboids for $a=0.3$, (b) $a = 1.0$, and (c) $a=2.0$.}
\label{fig:examples}
\end{figure}

\begin{figure}[htb]
\centerline{%
\includegraphics[width=0.7\columnwidth]{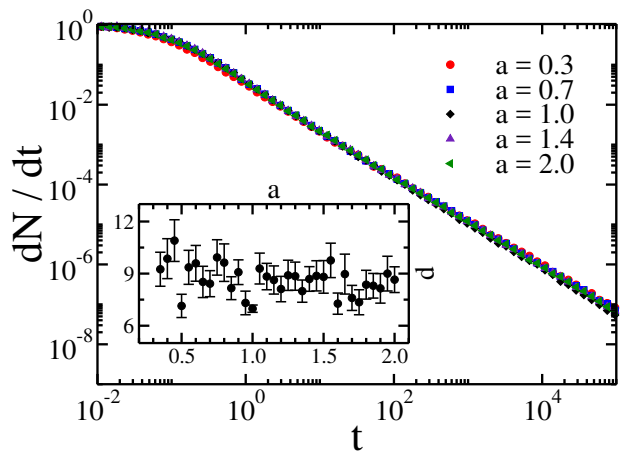}
}
\caption{Dependence of increments of the mean number of particles in a packing on dimensionless time for several values of the parameter $a$. Inset shows dependence of the exponent $d$ on $a$.}
\label{fig:d_a}
\end{figure}

Example packings are shown in Fig.\ref{fig:examples}.
As packing generation ends after a fixed number of iterations, there is no guarantee that obtained packings are saturated. Therefore, to estimate density of cuboids at saturation the knowledge of packing growth kinetics is necessary. It has been shown that for spherically symmetric particles, near saturation, packing fraction depends on dimensionless time according to the power-law \cite{Swendsen1981, Pomeau1980}:
\begin{equation}
\theta - \theta(t) =  A t ^{-\frac{1}{d}},
\label{fl}
\end{equation}
where $\theta$ is the saturated packing fraction and $A$ is a positive constant. Parameter $d$ is equal to the packing dimension. Relation (\ref{fl}) was confirmed numerically for $d \le 8$ \cite{Zhang2013, Ciesla2012, Ciesla2013frac}. For anisotropic particles, the power-law, in general, is still valid; however, exponent $d$ is then interpreted as the number of a particle's degrees of freedom \cite{Hinrichsen1986, Ciesla2013}. For some specific systems, the power-law is not valid \cite{Privman1991, Verma2018, Baule2017}. In case of cuboids with a square base, the power law seems to be valid (see Fig.\ref{fig:d_a}).
Parameter $d$ was estimated by fitting a power law $dN(t)/dt = A t^{-1/d - 1}$, which is a direct consequence of (\ref{theta}) and (\ref{fl}), to simulation data obtained for $t>1000$. Its error was calculated using exact differential method. 

Having determined $d$, $y=t^{-1/d}$ can be substituted and relation (\ref{fl}) rewritten in the form $\theta(y) = \theta - Ay$. Then, linear fit to the simulation data $(\theta(y), y)$ allows to determine saturated packing fraction. The errors of determined packing fractions are mainly due to inaccuracy of $d$ and are equal, at average, to $0.0023$. Analysis presented in \cite{Ciesla2016errors} and \cite{ciesla2017errors} suggests that the influence of limited number of RSA iterations and finite size effects are at least one order of magnitude smaller.

\begin{figure}[htb]
\centerline{%
\includegraphics[width=0.7\columnwidth]{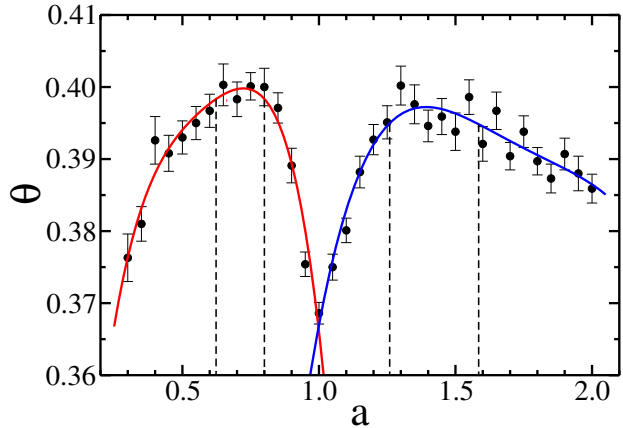}
}
\caption{The dependence of packing fraction on cuboid shape. Dots are simulation data and solid lines are polynomial fits: $\theta = 0.23796 + 0.92205 \, a - 2.165  \, a^2 + 2.4099 \, a^3 - 1.0388 \, a^4$ and $\theta = -0.8536 + 2.9388 \, a - 2.5582 \, a^2 + 0.98047 \, a^3 - 0.14053 \, a^4$ for $a<1$ and $a>1$, respectively. Dashed lines correspond to maximum width determined by the error of $\theta_{max}$.}
\label{fig:q_a}
\end{figure}

The dependence of estimated value of saturated packing fraction on the parameter $a$ is shown in Fig.\ref{fig:q_a}.
For cubes, the most symmetric shape, a local minimum is observed, which agrees with theoretical arguments \cite{Baule2013}. Similarly as for two dimensional shapes \cite{Vigil1989, Ciesla2015, Ciesla2016} and ellipsoids \cite{Sherwood1997}, slight anisotropy of packed shapes occurs in  denser packings. To estimate maxima positions we fitted 4-th order polynomial to the simulation data using the least square method. Maxima of these polynomials are: $\theta^1_{max} = 0.400$ and $\theta^2_{max} = 0.397$ for $a=0.72$ and $1.39$, respectively. As discussed earlier, errors of obtained saturated packing fractions equal $0.0023$. The error of parameter $a$ value, for which the maximum is reached, corresponds to the maximum width at the level defined by its error (see dashed lines in Fig.\ref{fig:q_a}). In the studied case, it gives intervals $[0.63, 0.80]$ for $\theta^1_{max}$ and $[1.26, 1.59]$ for $\theta^2_{max}$. Note that anisotropies giving the highest packing fractions are, in the range of errors, the same as for ellipsoids \cite{Sherwood1997}.
\section{Summary}
The packing density of cuboids with a square base arranged by random sequential adsorption is the highest for anisotropy within the range of $[0.63, 0.80]$ and $[1.26, 1.59]$ and equals to $0.400 \pm 0.002$, which is approximately 8\% higher as for cubes. Kinetics of packing growth obeys the power law (\ref{fl}) for all studied cases. The fastest cuboid-cuboid intersection detection test among all studied algorithms is based on separation axis theorem and it is faster than other, discussed here, methods by an order of magnitude.
\section*{Acknowledgements}
This work was supported by grant No. 2016/23/B/ST3/01145 of the National Science Centre, Poland. Numerical simulations were carried out with the support of the Interdisciplinary Centre for Mathematical and Computational Modelling (ICM) at University of Warsaw under grant no.\ G-27-8.
\section*{Appendix}
The implementation of the optimised SAT cuboid-cuboid intersection test in C++11:
\lstset{language=C++,
                basicstyle=\tiny\ttfamily,
                keywordstyle=\color{blue}\ttfamily,
                stringstyle=\color{red}\ttfamily,
                commentstyle=\color{green}\ttfamily,
                morecomment=[l][\color{magenta}]{\#}
}
\begin{lstlisting}
#include <cmath>
#include <array>

class Vector
{
private:
    std::array<double, 3> arr;
    
public:
    Vector() = default;
    Vector(const std::array<double, 3> &arr) : arr(arr) {}
    
    Vector rotate(double angleX, double angleY, double angleZ) const;
    Vector operator+(const Vector &other) const;
    Vector operator-(const Vector &other) const;
    double operator*(const Vector &other) const;
};

/* Counter-clockwise rotation around x, y and z axis, in mentioned order */
Vector Vector::rotate(double angleX, double angleY, double angleZ) const {
    double sin_ax = sin(angleX);
    double sin_ay = sin(angleY);
    double sin_az = sin(angleZ);
    double cos_ax = cos(angleX);
    double cos_ay = cos(angleY);
    double cos_az = cos(angleZ);
    
    std::array<double, 3> vectorOut;
    vectorOut[0] = arr[0] * (cos_ay * cos_az)
                 + arr[1] * (sin_ax * sin_ay * cos_az - cos_ax * sin_az)
                 + arr[2] * (cos_ax * sin_ay * cos_az + sin_ax * sin_az);
    vectorOut[1] = arr[0] * (cos_ay * sin_az)
                 + arr[1] * (sin_ax * sin_ay * sin_az + cos_ax * cos_az)
                 + arr[2] * (cos_ax * sin_ay * sin_az - sin_ax * cos_az);
    vectorOut[2] = arr[0] * (-sin_ay)
                 + arr[1] * (sin_ax * cos_ay)
                 + arr[2] * (cos_ax * cos_ay);
    return Vector{vectorOut};
}

Vector Vector::operator+(const Vector &other) const {
    return Vector{{arr[0] + other.arr[0], arr[1] + other.arr[1], arr[2] + other.arr[2]}};
}

Vector Vector::operator-(const Vector &other) const {
    return Vector{{arr[0] - other.arr[0], arr[1] - other.arr[1], arr[2] - other.arr[2]}};
}

double Vector::operator*(const Vector &other) const {
    return arr[0] * other.arr[0] + arr[1] * other.arr[1] + arr[2] * other.arr[2];
}


struct Cuboid {
    std::array<double, 3> halfSize;
    Vector position;
    Vector normalAxes[3];
    
    Cuboid(Vector position, const std::array<double, 3> &halfSize,
           double angleX, double angleY, double angleZ);
};

Cuboid::Cuboid(Vector position, const std::array<double, 3> &halfSize,
               double angleX, double angleY, double angleZ)
: halfSize(halfSize), position(position) {
    normalAxes[0] = Vector{{1, 0, 0}}.rotate(angleX, angleY, angleZ);
    normalAxes[1] = Vector{{0, 1, 0}}.rotate(angleX, angleY, angleZ);
    normalAxes[2] = Vector{{0, 0, 1}}.rotate(angleX, angleY, angleZ);
}

/* Checks if a separating axis exists, returns true when it was not found, so cuboids do intersect */
bool overlap(Cuboid *cube1, Cuboid *cube2) {
    Vector T = cube2->position - cube1->position;

    double WA = cube1->halfSize[0];
    double HA = cube1->halfSize[1];
    double DA = cube1->halfSize[2];
    double WB = cube2->halfSize[0];
    double HB = cube2->halfSize[1];
    double DB = cube2->halfSize[2];

    Vector Ax = cube1->normalAxes[0];
    Vector Ay = cube1->normalAxes[1];
    Vector Az = cube1->normalAxes[2];
    Vector Bx = cube2->normalAxes[0];
    Vector By = cube2->normalAxes[1];
    Vector Bz = cube2->normalAxes[2];
    
    double Rxx = Ax * Bx;   double Rxy = Ax * By;   double Rxz = Ax * Bz;
    double Ryx = Ay * Bx;   double Ryy = Ay * By;   double Ryz = Ay * Bz;
    double Rzx = Az * Bx;   double Rzy = Az * By;   double Rzz = Az * Bz;

    if (abs(T*Ax) > WA + abs(WB*Rxx) + abs(HB*Rxy) + abs(DB*Rxz))
        return false;
    else if (abs(T*Ay) > HA + abs(WB*Ryx) + abs(HB*Ryy) + abs(DB*Ryz))
        return false;
    else if (abs(T*Az) > DA + abs(WB*Rzx) + abs(HB*Rzy) + abs(DB*Rzz))
        return false;
    else if (abs(T*Bx) > abs(WA*Rxx) + abs(HA*Ryx) + abs(DA*Rzx) + WB)
        return false;
    else if (abs(T*By) > abs(WA*Rxy) + abs(HA*Ryy) + abs(DA*Rzy) + HB)
        return false;
    else if (abs(T*Bz) > abs(WA*Rxz) + abs(HA*Ryz) + abs(DA*Rzz) + DB)
        return false;
    else if (abs(T*Az*Ryx - T*Ay*Rzx) > abs(HA*Rzx) + abs(DA*Ryx) + abs(HB*Rxz) + abs(DB*Rxy))
        return false;
    else if (abs(T*Az*Ryy - T*Ay*Rzy) > abs(HA*Rzy) + abs(DA*Ryy) + abs(WB*Rxz) + abs(DB*Rxx))
        return false;
    else if (abs(T*Az*Ryz - T*Ay*Rzz) > abs(HA*Rzz) + abs(DA*Ryz) + abs(WB*Rxy) + abs(HB*Rxx))
        return false;
    else if (abs(T*Ax*Rzx - T*Az*Rxx) > abs(WA*Rzx) + abs(DA*Rxx) + abs(HB*Ryz) + abs(DB*Ryy))
        return false;
    else if (abs(T*Ax*Rzy - T*Az*Rxy) > abs(WA*Rzy) + abs(DA*Rxy) + abs(WB*Ryz) + abs(DB*Ryx))
        return false;
    else if (abs(T*Ax*Rzz - T*Az*Rxz) > abs(WA*Rzz) + abs(DA*Rxz) + abs(WB*Ryy) + abs(HB*Ryx))
        return false;
    else if (abs(T*Ay*Rxx - T*Ax*Ryx) > abs(WA*Ryx) + abs(HA*Rxx) + abs(HB*Rzz) + abs(DB*Rzy))
        return false;
    else if (abs(T*Ay*Rxy - T*Ax*Ryy) > abs(WA*Ryy) + abs(HA*Rxy) + abs(WB*Rzz) + abs(DB*Rzx))
        return false;
    else if (abs(T*Ay*Rxz - T*Ax*Ryz) > abs(WA*Ryz) + abs(HA*Rxz) + abs(WB*Rzy) + abs(HB*Rzx))
        return false;

    return true;
}
\end{lstlisting}

\bibliographystyle{aip}
\bibliography{main}


\end{document}